\documentclass[twocolumn]{svjour3}
\usepackage{amsmath}
\usepackage{graphicx}
\begin{document}
\title{Dense granular flow at the critical state: 
       maximum entropy and topological disorder}
\author{Matthew R. Kuhn}%
\institute{Donald P. Shiley School of Engrg., Univ. of Portland, 
           Portland, OR 97203. \email{kuhn@up.edu}.  Tel. 1-503-943-7361.
           Fax. 1-503-943-7316.}
%
%
\titlerunning{Critical state flow as a disordered process}
\maketitle
\begin{abstract}
After extensive quasi-static shearing, dense dry granular flows 
attain a steady-state condition of porosity and deviatoric stress,
even as particles are continually rearranged.
The Paper considers two-di\-men\-sional flow and derives
the probability distributions of two topological measures
of particle arrangement ---
coordination number and void valence ---
that maximize topological entropy.
By only considering topological dispersion, the method
closely predicts the distribution of void valences,
as measured in discrete element (DEM) simulations.
Distributions of coordination number are also derived
by considering packings that are geometrically and kinetically
consistent with the particle sizes and friction coefficient.
A cross-entropy principle results in 
a distribution of coordination numbers that closely fits
DEM simulations.
\end{abstract}
\section{Introduction}
The critical state principle, a unifying concept in geomechanics,
holds that dense confined granular materials attain a steady-state
condition of flow after extensive slow shearing\ \cite{Schofield:1968a}. 
A distinctive feature of such flow is the continual micro-scale rearrangement
of particles even as bulk characteristics~--- stress, density,
and fabric~--- remain nearly constant. 
Micro-scale rearrangements
are expressed in three ways: (a)~statically (kinetically), as alterations of
inter-particle contact forces, (b)~geometrically, as changes
in the particles' positions or in the local density,
and (c)~topologically, as changes in the load-bearing contact network
among the particles. 
This Paper addresses the latter form of granular
arrangement as expressed in a two-dimensional (2D) setting,
focusing on the local coordination number
(number of contacts per particle) and the local
void valence (number of particles surrounding a void).
The analysis applies to the critical state flow of
frictional materials of sufficient density
to develop a load-bearing network of contacts 
during slow (quasi-static) shearing.
\par
\begin{figure}
  \centering
  \includegraphics{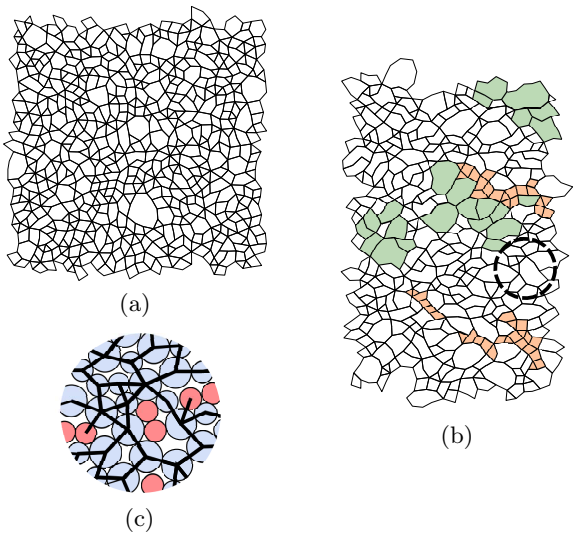}
  \caption{Particle graphs of disk assemblies: 
           (a)~initial state, (b)~steady-state
           flow in biaxial compression, 
           (c)~detail. 
           In the electronic version, rattler particles are shown in red,
           and clusters of small and large voids are highlighted
           in orange and green, respectively.}
  \label{fig:Particle-graphs}
\end{figure}

The discrete 2D topological arrangement of load-bearing particles is
conveniently represented with a planar \emph{particle graph},
a tessellation in which edges,
nodes, and faces represent inter-particle contacts, particles,
and voids, respectively\ \cite{Satake:1992a}.
Fig.~\ref{fig:Particle-graphs} shows particle graphs from discrete
element (DEM) simulations\ \cite{Cundall:1979a} 
of a two-di\-men\-sion\-al assembly of
circular disks at two states: an initial state of
676 densely arranged disks 
and a deformed state in which the assembly width has been slowly reduced
by a horizontal strain $\varepsilon_{11}=-20\%$ 
while maintaining the original, constant mean stress
~$p=\frac{1}{2}(\sigma_{11}+\sigma_{22})$.
Such biaxial loading departs from the initial isotropic condition,
producing deviatoric stress $\sigma_{11}-\sigma_{22}$ and 
inducing steady-state (critical state) flow in
which continued, plastic deformation progresses at constant volume,
mean stress, and deviatoric stress\ \cite{Schofield:1968a}. 
Gross topological parameters, such as the
average coordination number and average void valence
(i.e., the number of sides of an $l$-polygon void), also remain
constant during such flow\ \cite{Thornton:2000a,Pena:2009a}. 
The micro-scale topology, however, is continually altered, to the
extent that a further small deformation of the assembly in 
Fig.~\ref{fig:Particle-graphs}(b)
of as little as $\Delta\varepsilon_{11}=2\%$ will produce a particle
graph that is scarcely recognizable from the one shown. 
During deformation,
an existing polygonal void
can split into two smaller voids when fresh contacts
are newly established among 
unloaded, \emph{rattler}
particles within the original void
(i.e., particles with zero or one contact)\ \cite{Kruyt:2012a}. 
Likewise, adjacent voids can merge when contacts disengage
along shared edges. 
These transmutations, similar to those in hexagonal cellular and
froth structures\ \cite{Weaire:1984a}, produce a disordered and constantly 
changing (and seemingly random) topology.
\par 
We view this continual transmutation of particle arrangement as a
maximally disordered process, in which void polygons split and merge
in a random, disordered manner, subject only to bulk
topological constraints. 
The principle of maximum entropy has been applied
to granular materials for several decades.
In these studies, disorder is usually expressed as the
Shannon entropy\ \cite{Shannon:1948a,BenNaim:2008a}.
Brown et al.\ \cite{Brown:2000b} conducted
experiments on two-dimensional assemblies of spheres, 
and by applying
a back-and-forth shearing,
disorder in the local density and coordination
number increased with each shearing cycle. 
The Jaynes\ \cite{Jaynes:1957a} maximum entropy (MaxEnt) formalism
has typically been used to solve the condition of maximum disorder.
This approach has been applied 
to the local fabric of disk assemblies by categorizing 
voids into several canonical types\ \cite{Shahinpoor:1983b,Brown:2000a}.
Similar maximum entropy approaches have also been applied to
the local packing 
density\ \cite{Moroto:1983a,Edwards:1989a,Kumar:2005a,Yoon:2012a},
to contact forces\ \cite{Coppersmith:1996a,Edwards:2001a,Chakraborty:2010a},
to contact orientations\ \cite{Troadec:2002a},
and to contact displacements and bulk elastic moduli\ \cite{Rothenburg:2009a}.
\par
The Euler equation, $M=N+L-1$, 
applies to the bulk topological quantities of a particle graph:
$M$ is the number of contacts (edges), $N$ is the
number of particles (nodes), and $L$ is the number of
void polygons (faces).
The equation holds for any connected planar graph, 
but we restrict attention to the \emph{load-bearing} 
subgraph of an assembly, which excludes peninsular and
island particles (rattlers), such as those apparent
in Fig.~\ref{fig:Particle-graphs}(c). 
(Such particles, shown in red,
are typically ``nudged'' along by their neighbors
until the surrounding void eventually collapse onto them.)
For large assemblies,
the average coordination
number $\overline{n}$ among load-bearing particles and the average
void valence $\overline{l}$ are defined as 
\begin{equation}
  \overline{n}=2M/N\:,\quad
  \overline{l}= 2M/L=
  2\overline{n}/(\overline{n}-2)
  \label{eq:CoordValence}
\end{equation}
which will serve as the constraints on possible
particle graphs. 
The corresponding micro-scale quantities are the
number of contacts $n_{j}$ of the ``$j$''th load-bearing particle
and the valence $l_{i}$ of the ``$i$''th $l$-polygon void.
\par
We develop three methods for estimating the probability distributions
of $n_{j}$ and $l_{i}$.
In the next section, a maximum entropy principle (MaxEnt) 
is used to maximize the disorder of the two distributions.
Although this approach leads to a reasonable approximation of the void
valence $l_{i}$ distribution, the estimated distribution of coordination
numbers $n_{j}$ misses many features observed in discrete element (DEM)
simulations.
In Section~\ref{sec:ModelII}, 
we invoke certain geometric and kinetic constraints to
develop a second distribution of coordination numbers.
This distribution is an improvement over the first, and it also resolves
the effect of inter-granular friction on the coordination number.
In Section~\ref{sec:ModelIII}, the two methods are combined with a minimum
cross-entropy (MinXEnt) principle, which yields
an improved estimate of the coordination number distribution.
\section{Maximum disorder theory (Model~I)}\label{sec:ModelI}
We first develop a model, based upon a maximum disorder principle,
for estimating the probabilities $P_{l}^{\text{I}}$ and $P_{n}^{\text{I}}$
of encountering a void with valence $l$ and a particle with coordination
number $n$. 
The superscript ``I'' denotes probabilities derived from
this first approach.
We begin with a method of 
constructing a particle graph 
from a journal $\{s_{i}\}$ of integer pairs $s_{i}=(l_{i},\Delta M_{i})$, a
journal that represents a single micro-state of topology. 
Each journal entry $s_{i}$ corresponds to a single ``$i$th'' void
polygon (face) of valence $l_{i}$ that has been appended to
the particle graph by adding $\Delta M_{i}$ new contacts (edges).
After describing the construction of a journal,
we then establish
constraints on such journals. 
Assuming that all journals that meet
these constraints are equiprobable 
during steady-state flow 
and that the most likely topology maximizes the disorder
of the system, we derive the expected probabilities of
pairs $(l,\Delta M)$ and, by extension, the probabilities $P_{l}^{\text{I}}$
of valences $l$. 
A duality principle leads to the complementary probabilities
of coordination numbers, $P_{n}^{\text{I}}$. 
These expected probabilities,
$P_{n}^{\text{I}}$ and $P_{l}^{\text{I}}$, 
are then compared with those measured in DEM simulations.
\par
The disorder of a particle graph is characterized by the
number of ways in which similar, equiprobable graphs can be constructed
from sets of polygonal faces. 
%
In this regard, we start with a scheme for identifying
and ``counting'' these graphs (i.e., micro-states),
each micro-state being a nearly random sequence $\{s_i \}$
of pairs $s_i = (l_{i}, \Delta M_{i})$.
\par
\begin{figure}
  \centering
  \includegraphics{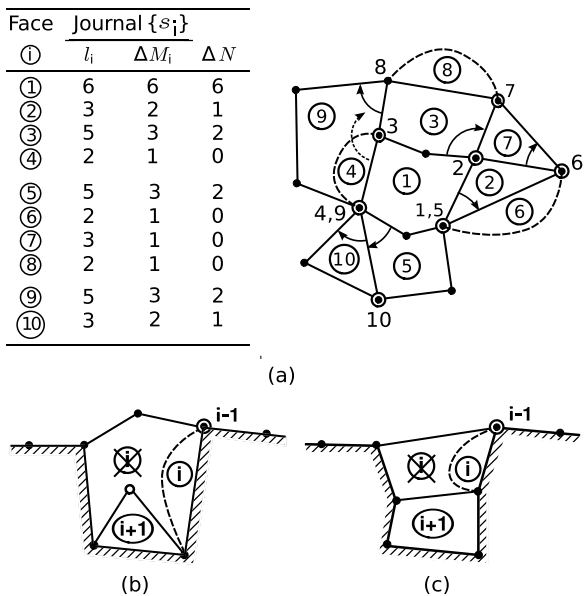}
  \caption{Constructing a particle graph from journal $\left\{ s_{i}\right\}$.
           (a)~Seven-face graph with three pseudo-faces. 
           (b)~and (c)~Two cases in which a pseudo-face is required.
           Hatching indicates the perimeter of the existing graph.}
  \label{fig:Journal}
\end{figure}

We now describe the meaning of such journals
(micro-states), referring to 
Fig.~\ref{fig:Journal}a and the construction of this
seven-polygon graph.
Starting from a seed polygon and one of its vertices
(labeled \raisebox{0.5pt}{\textcircled{\raisebox{-.7pt} {\small 1}}}
and ``1'' in Fig.~\ref{fig:Journal}),
a planar graph can be constructed by progressively appending faces to
the graph.
Each new face of valence $l_{i}$ adds $\Delta M_{i}$ edges and
$\Delta N_{i}$ vertices ($\Delta N_{i}=\Delta M_{i}-1$).
A journal $\left\{ s_{i}\right\}$ of integer pairs
$s_{i}=(l_{i},\Delta M_{i})$ records this process
(Fig.~\ref{fig:Journal}a).
%
Each new face \raisebox{0.5pt}{\textcircled{\raisebox{-.7pt} {\small i}}}
begins from the terminal vertex ``$i-1$'' of the previous face. 
In this manner, new faces, having pairs $(l_{i},\Delta M_{i})$,
are spiraled counter-clockwise around 
the perimeter of the existing graph,
with each \raisebox{0.5pt}{\textcircled{\raisebox{-.7pt} {\small i}}}th face 
sharing $l_{i}-\Delta M_{i}$ edges with the existing graph. 
The curved arrows in Fig.~\ref{fig:Journal}a
show that each added face, beginning from the previous $i-1$ terminus,
includes the edge of the existing perimeter that lies in front
of (i.e. counter-clockwise relative to)
the previous terminus, and the new face also
includes the edge immediately clockwise around the terminus node.
For example, 
face \raisebox{0.5pt}{\textcircled{\raisebox{-.7pt} {\small 3}}},
rather than the later
face \raisebox{0.5pt}{\textcircled{\raisebox{-.7pt} {\small 7}}},
is added from terminus~2, even though both faces share this node.
Several faces can start from the same
node, whereas some nodes are not the start of any face.
\par
A unique graph should result from any given journal, and a unique
journal should be associated with any given graph in which
the seed has been specified.
To achieve these qualities,
we require three additional conditions with each added face.
\begin{enumerate}
\item
No edge of a new face 
\raisebox{0.5pt}{\textcircled{\raisebox{-.7pt} {\small i}}}
should lie behind (i.e. clockwise relative to) the $i-1$ terminus along
the existing perimeter: all edges of
\raisebox{0.5pt}{\textcircled{\raisebox{-.7pt} {\small i}}}
should be new edges or should be existing edges that lie in front
of the $i-1$ node.
With some nodes, a pseudo-face with two edges can be used
to skip across an existing perimeter edge.
For example, starting at node~3, 
the face \raisebox{0.5pt}{\textcircled{\raisebox{-.7pt} {\small 9}}}
could not be added as the fourth face, since this face would include
an edge (labeled 3--8) that lies behind node~3.
The pseudo-face \raisebox{0.5pt}{\textcircled{\raisebox{-.7pt} {\small 4}}}
skips to the next node, labeled~4,
from which face \raisebox{0.5pt}{\textcircled{\raisebox{-.7pt} {\small 5}}}
emanates.
\item
When considering the ordered list of all nodes of a new face
\raisebox{0.5pt}{\textcircled{\raisebox{-.7pt} {\small i}}},
beginning with the previous $i-1$ terminus and moving clockwise
around the face,
these nodes should consist of existing (perimeter) nodes followed
by added nodes.
A counter-example is shown in
Fig.~\ref{fig:Journal}b, where the six nodes (solid dots) of an improper face
\raisebox{0.5pt}{\textcircled{\raisebox{-.7pt} {\small i}}} (crossed)
have the pattern T-T-F-T-T-T ($\text{T}=\text{ }$existing,
$\text{F}=\text{ }$new).
If this face in Fig.~\ref{fig:Journal}b
was added to the graph, 
the interior triangular face would be stranded and could not later
be added to the journal.
A proper pseudo-face \raisebox{0.5pt}{\textcircled{\raisebox{-.7pt} {\small i}}}
(not crossed) must be used to skip to its terminus, after which the
triangular face $i+1$ would be added.
The improper face
\raisebox{0.5pt}{\textcircled{\raisebox{-.7pt} {\small i}}}
(crossed) could be added later, after the process has spiraled fully around the
graph once again.
\item
When considering the ordered list of nodes in the existing perimeter, 
beginning with the previous terminus $i-1$
and moving counter-clockwise around the perimeter, these nodes
should consist of nodes that are part of the new face
\raisebox{0.5pt}{\textcircled{\raisebox{-.7pt} {\small i}}}
followed by nodes that are not in
\raisebox{0.5pt}{\textcircled{\raisebox{-.7pt} {\small i}}}.
A counter-example is shown in Fig.~\ref{fig:Journal}c,
with an improper face
\raisebox{0.5pt}{\textcircled{\raisebox{-.7pt} {\small i}}} (crossed).
Beginning with the node $i-1$, we have the pattern
T-T-F-F-T-T-T-T-$\ldots$ along the existing perimeter
($\text{T}=\text{ }$included in the nodes of 
\raisebox{0.5pt}{\textcircled{\raisebox{-.7pt} {\small i}}},
$\text{F}=\text{ }$not included in the nodes of
\raisebox{0.5pt}{\textcircled{\raisebox{-.7pt} {\small i}}}),
where the two bottom-most nodes are not part of the
improper face 
\raisebox{0.5pt}{\textcircled{\raisebox{-.7pt} {\small i}}} (crossed).
To avoid stranding the bottom
quadrilateral face, a pseudo-face
\raisebox{0.5pt}{\textcircled{\raisebox{-.7pt} {\small i}}}
skips ahead so that the proper bottom face 
$i+1$ (circled) is added to the graph.
%
\end{enumerate}
The journal of a given graph can be efficiently
constructed when the graph is represented with the data structure
of a doubly-connected edge list (DCEL)\ \cite{Preparata:1985a}, which permits
rapid counter-clockwise traversals of the faces and nodes.
\par
Every journal $\left\{ s_{i}\right\}$
corresponds to a unique planar graph (a topological
micro-state), although the same graph can be constructed from different
journals, depending upon the chosen seed.
The total number
of faces, $L$, in the load-bearing graph is equal to the numbers of
pairs in the journal, excluding those 
of valence~2.
The set of all possible journals of length $L$ comprises a topological
configuration space. 
We require, however, that a journal $\left\{ s_{i}\right\}$
be consistent with the 
known coordination number $\overline{n}$ and valence $\overline{l}$.
The total number of contacts (edges) and particles (vertices) are
\begin{equation}
  M=\sum_{i=1}^{L}\Delta M_{i}\:,\quad 
  N=1+\sum_{i=1}^{L}\Delta N_{i}=1+\sum_{i=1}^{L}(\Delta M_{i}-1)
  \label{eq:MDeltaN}
\end{equation}
excluding faces of valence 1 or 2.
Substituting Eqs.~(\ref{eq:CoordValence}$_{1}$) and~(\ref{eq:MDeltaN}$_{1}$)
into the Euler equation,
dividing by $L$, and neglecting the insignificant $1/L$ yields a constraint
on the journal $\left\{ s_{i}\right\} $, 
\begin{equation}
  \left\langle\Delta M_{i}\right\rangle 
  \equiv \frac{1}{L}\sum_{i=1}^{L}\Delta M_{i}=
  \frac{\overline{n}}{\overline{n}-2}
  \label{eq:avgMi}
\end{equation}
The journal must also be consistent with the 
average valence $\overline{l}$ [Eq.~(\ref{eq:CoordValence}$_{2}$)]:
\begin{equation}
  \left\langle l_{i}\right\rangle \equiv
  \frac{1}{L}\sum_{i=1}^{L}l_{i}=\overline{l}=
  \frac{2\overline{n}}{\overline{n}-2}
  \label{eq:avgli}
\end{equation}
\par
Elements in a journal $\left\{ s_{i}\right\} $ correspond to
the faces (voids) of a particle graph. 
The journals of small assemblies can be affected
by the choice of seed,
a matter addressed below.
In the limit of large assemblies,
we can assign discrete probabilities $P_{l,\Delta M}^{\text{I}}$
to the likelihood of each ``$l$ species'' and its ``$\Delta M$ sub-species''
among the $L$ faces.
The random variables $l$ and $\Delta M$ are assumed to be independent.
Eq.~(\ref{eq:avgMi})
implies a constraint on these probabilities:
\begin{equation}
  \left\langle \Delta M\right\rangle \equiv
  \sum_{l=3}^{\infty}\sum_{\Delta M=1}^{l-1}\Delta M\, P_{l,\Delta M}^{\text{I}}=
  \frac{\overline{n}}{\overline{n}-2}
  \label{eq:Constraint2}
\end{equation}
The outer sum explicitly excludes valences 1 and 2, as these
values apply only to non-load-bearing particles, such as peninsular
particles (rattlers).
The inner sum ignores pairs with $\Delta M=l$, as this
situation arises only with the initial, seed face.
A second constraint results from the average valence $\overline{l}$
of an assembly (Eq.~(\ref{eq:avgli})):
\begin{equation}
  \left\langle l\right\rangle \equiv
  \sum_{l=3}^{\infty}\sum_{\Delta M=1}^{l-1}l\, P_{l,\Delta M}^{\text{I}}=
  \frac{2\overline{n}}{\overline{n}-2}
  \label{eq:Constraint3}
\end{equation}
\par
The full mechanical description of an assembly's incremental evolution
requires solution of an $N$-body
problem, involving $3N$ equilibrium equations, 
while applying $2M$ contact elastic-frictional rules and any applicable
boundary constraints~\cite{Kuhn:2005b,Agnolin:2007c}.
Although this non-linear problem is solvable, 
its solution also requires full information
about the initial particle positions and contact forces. 
Lacking such information, 
we simply estimate the
probabilities $P_{l,\Delta M}^{\text{I}}$ that describe the most
likely topological condition.
The most likely probability set~--- the set that optimally
respects the missing information and, hence, corresponds to the greatest
number of similar journals within the configuration space~--- is
the one which maximizes the topological disorder 
(Shannon entropy)\ \cite{Jaynes:1957a,BenNaim:2008a},
\begin{equation}
  H_{l}^{\text{I}}=
  -\sum_{l=3}^{\infty}\sum_{\Delta M=1}^{l-1}
     P_{l,\Delta M}^{\text{I}}\ln\left(P_{l,\Delta M}^{\text{I}}\right)
  \label{eq:Hl}
\end{equation}
such that $\partial H_{l}^{\text{I}} / \partial P_{l,\Delta M}^{\text{I}} = 0$,
while satisfying the moment constraints of Eqs.~(\ref{eq:Constraint2})
and~(\ref{eq:Constraint3}).
By applying the Jaynes formalism\ \cite{Jaynes:1957a} 
in maximizing $H_{l}^{\text{I}}$ (i.e., using the maximum entropy ``MaxEnt''
principle), 
the probabilities $P_{l,\Delta M}^{\text{I}}$ and partition function $Z_{l}^{\text{I}}$ are
\begin{gather}
  P_{l,\Delta M}^{\text{I}}=
  \frac{1}{Z_{l}^{\text{I}}\left(\lambda_{l,1},\lambda_{l,2}\right)}
  \exp\left(-\lambda_{l,1}l-\lambda_{l,2}\Delta M\right)
  \label{eq:PlDeltaM}\\
  Z_{l}^{\text{I}}\left(\lambda_{l,1},\lambda_{l,2}\right)=
  \sum_{l=3}^{\infty}\sum_{\Delta M=1}^{l-1}
  \exp\left(-\lambda_{l,1}l-\lambda_{l,2}\Delta M\right)
  \label{eq:Partition}
\end{gather}
with two Lagrange multipliers, $\lambda_{l,1}$ and $\lambda_{l,2}$, that
satisfy Eqs.~(\ref{eq:Constraint2}) 
and~(\ref{eq:Constraint3}).
The probability $P_{l}^{\text{I}}$ of encountering a void of valence
$l$ is the exponential marginal probability
\begin{equation}
  P_{l}^{\text{I}}=\sum_{\Delta M=1}^{l-1}P_{l,\Delta M}^{\text{I}}
  \label{eq:PastlDeltaM}
\end{equation}
\par
The probabilities of coordination numbers $n$
are developed with the dual of the particle graph
(the \emph{void graph}),
in which particles and voids assume the complementary
roles of faces and vertices~\cite{Satake:1992a}. 
The alternative journal 
$\{s_{j}^{\prime}\}=\{(n_{j},\Delta M_{j})\}$
is constructed by adding particles
(now faces in the dual, void graph) of coordination
number $n_{j}$ 
around a seed void (now a vertex). 
In this manner, we can develop complementary
statistics on $n$ 
(see Table~\ref{tab:Correspondence}).
\begin{table}
  \caption{Duality of the valence and coordination number
           problems\label{tab:Correspondence}}
  \begin{tabular}{cc}
    \hline
    Void valence & Particle coordination number\\
    \hline 
    particles, contacts, voids & voids, contacts, particles\\
    $\{s_{i}\}=\{(l_{i},\Delta M_{i})\}$ & 
      $\{s_{i}^{\prime}\}=\{(n_{i},\Delta M_{i})\}$\\
   $l_{i}\in\{3,4,\ldots\}$ & 
      $n_{i}\in\{2,3,\ldots,n_{\text{max}}\}$\\
   $\Delta M_{i}\in\{1,2,\ldots,l_{i}-1\}$ & 
      $\Delta M_{i}\in\{1,2,\ldots,n_{i}-1\}$\\
    $P_{l,\Delta M}^{\text{I}}$ & $P_{n,\Delta M}^{\text{I}}$\\
    Eq.~(\ref{eq:avgMi}) &
      ${\displaystyle \left\langle \Delta M\right\rangle =
        \overline{n}/2}$
        \\
    Eq.~(\ref{eq:avgli}) &
      ${\displaystyle \left\langle n\right\rangle =
        \overline{n}}$\\
    \hline
  \end{tabular}
\end{table}
This approach differs in three respects from the previous analysis
of void valence $l$. 
First,
the minimum coordination number is 2, with $n_{i}\in\{2,3,\ldots\}$,
since a particle can have two contacts, but particles
with fewer than two contacts cannot serve within 
a load-bearing contact network. 
Second, among the faces (particles) with $n_{i}=2$, some
are truly particles with two contacts; others can be pseudo-faces.
A list must be maintained to distinguish between these two cases.
Finally, the maximum coordination
number $n_{\text{max}}$ is limited by geometric exclusion and will depend
upon the range of particle sizes, a matter more fully
developed in Section~\ref{sec:ModelII}.
\subsection{Simulation methods and results --- Method I}
\begin{figure}
  \centering
  \includegraphics{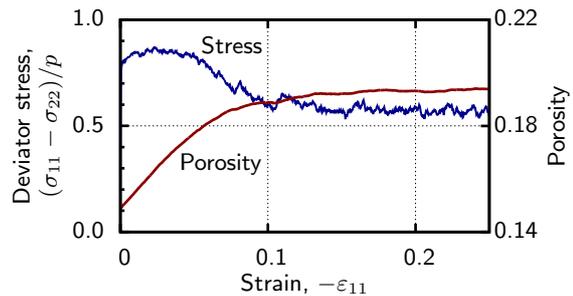}
  \caption{Stress and porosity from DEM simulations.}
  \label{fig:Stress-and-porosity}
\end{figure}

Statistics of valence and coordination number were measured in
assemblies of 676 bi-disperse disks that were sheared in biaxial compression
(Fig.~\ref{fig:Particle-graphs}). 
The two disk varieties have ratios
of 1.5:1 in size, 1:2.25 in number, and 1:1 in cumulative area
(that is, 468 particles of size $1.0$ and 208 of size $1.5$).
The assemblies were small enough to prevent gross non-homogeneity
in the form shear bands, 
yet large enough to capture the average, bulk material behavior.
To develop more robust statistics, 168 different assemblies were created
by compacting random sparse mixtures of the two disk sizes
into dense isotropic packings within periodic boundaries until
the average contact indentation was $0.0002$ times the average radius.
Linear contact stiffnesses were applied between particles 
with equal tangential and normal coefficients 
($k^{\text{t}}=k^{\text{n}}$), and
the friction coefficient $\mu=0.50$ was enforced during the pair-wise
particle interactions~\cite{daCruz:2005a}.
Using the discrete element (DEM) algorithm,
the initially square assemblies were horizontally 
compressed in increments $\Delta\epsilon_{11}=1\times10^{-6}$
while maintaining a constant mean stress of $2\times 10^{-4}k^{\text{n}}$.
Stress and volumetric behavior are shown in 
Fig.~\ref{fig:Stress-and-porosity}.
A large initial stiffness cause the deviatoric
stress to rise quickly from zero to a peak stress at
strain $-\varepsilon_{11}=2\%$.
The critical state condition was 
attained at compressive strains $-\varepsilon_{11}$ of 16--18\%.
During subsequent steady-state deformation, the particle graph of each
assembly was interrogated at five strains between $-\varepsilon_{11}=16\%$
and~$25\%$ (Fig.~\ref{fig:Stress-and-porosity}). 
Applying the ergodicity principle, micro-sate statistics were averaged
across the five strains and the 168 assemblies, involving 840 particle
graphs containing over 300,000 void faces. 
\par
At the critical state,
the average valence $\overline{l}=5.24$,
and the average coordination number $\overline{n}=3.23$ among load-bearing
particles (i.e., excluding rattlers). 
About 11\% of particles were non-load-bearing rattlers.
The probabilities $P_{l,\Delta M}^{\text{I}}$ are 
nearly the
same across all assemblies and at all strains,
although averages $\overline{l}$ and $\overline{n}$ will depend upon
the particular material properties (for example, a larger coefficient $\mu$
will increase $\overline{l}$\ \cite{Kruyt:2013a}).
\par
\begin{figure}
  \centering
  \includegraphics{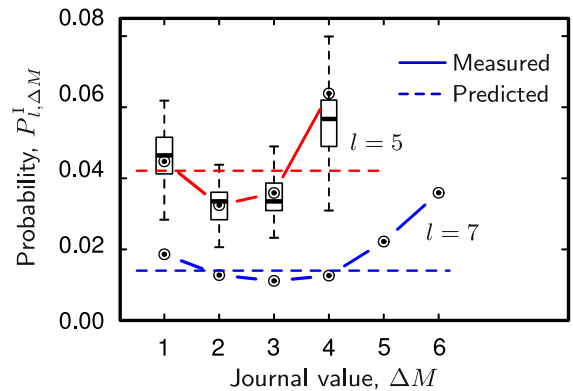}
  \caption{Probabilities $P_{l,\Delta M}^{\text{I}}$ 
           for void valences $l=5$ and~7
           among 840 particle graphs, compared with predictions
           of Model~I.
           Box-plots are for a single graph, but with probabilities
           generated among the choice of fifty seeds.}
  \label{fig:Probabilities-DM}
\end{figure}

Predictions of topological statistics
are based upon Eqs.~(\ref{eq:Constraint2})--(\ref{eq:PastlDeltaM})
and Table~\ref{tab:Correspondence}, using the measured
averages $\overline{l}$ and $\overline{n}$. 
An upper bound
of 14 was applied to $l$, as larger voids were not observed.
Micro-state (journal) statistics are shown in Fig.~\ref{fig:Probabilities-DM}
for two void valences, $l=5$ and~7. 
This figure was produced
by reconstructing the journals of 840 particle graphs.
Eqs.~(\ref{eq:Constraint2})--(\ref{eq:Partition}) predict varying probabilities
$P_{l,\Delta M}^{\text{I}}$ across valences $l$, but they predict a uniform
probability for each $\Delta M$ within a single valence~$l$. 
The latter prediction is not realized in the simulations
(Fig.~\ref{fig:Probabilities-DM}),
which exhibit a variation of $P_{l,\Delta M}^{\text{I}}$ with $\Delta M$. 
Additions $\Delta M$ of one and $l-1$ are more frequent
than those with $\Delta M$ near $l/2$.
This result 
suggests that a somewhat greater order
is realized in the simulations than predicted by the theory
(i.e., the assumption that $l$ and $\Delta M$ are independent
is not supported by the data).
The greater topological order is reflected in a measured entropy
$H_{l}^{\text{I}}$ of 3.191 
compared with a prediction of 3.306 (Eq.~\ref{eq:Hl}).
\par
The box-plot within 
Fig.~\ref{fig:Probabilities-DM} also shows the effect of the
choice of a seed on the statistics of voids having five edges ($l=5$).
A single graph was chosen among the 840 graphs, and fifty seeds were
applied.
Although the choice of seed does affect the distribution of $\Delta M$
values, the effect (expressed as a standard deviation) is less than
9\% of the mean probabilities, and the resulting entropy $H_{l}^{\text{I}}$
is affected by less than $0.5\%$.
\par
\begin{figure}
  \centering
  \includegraphics{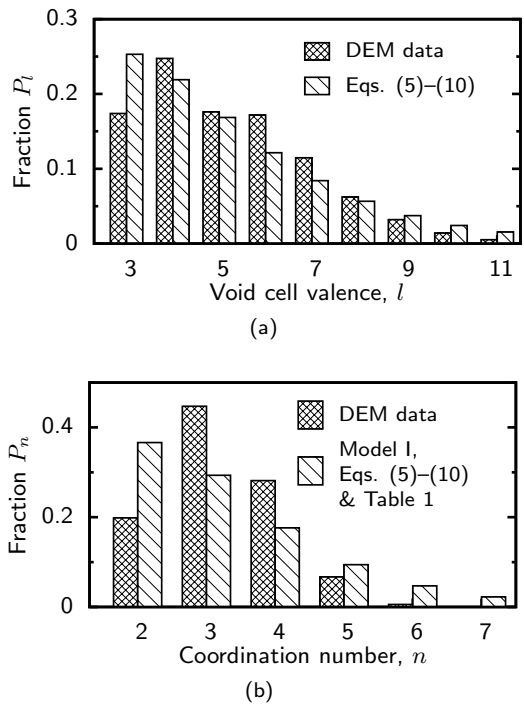}
  \caption{Estimated distributions 
           from the Model~I compared
           with DEM simulation results.}
  \label{fig:Estimated-and-measured}
\end{figure}

Fig.~\ref{fig:Estimated-and-measured}a shows the predicted and actual
probabilities $P_{l}^{\text{I}}$ of void valence,
based upon the average $\overline{l}$ (Eq.~\ref{eq:PastlDeltaM}). 
The valence distributions
are in general agreement, although the theory predicts a greater frequency
of triangular voids and of voids having valence greater than 9,
whereas the theory slightly under-predicts the frequencies of valences
4 through~8.
Experiments with assemblies having different friction
coefficients $\mu=0.1$--$0.7$ show similar features 
as Fig.~\ref{fig:Estimated-and-measured}a,
although $\overline{l}$ increases with $\mu$, and the histograms, both
measured and predicted, broaden to the right.
\par
A similar analysis of coordination number is based upon 
Table~\ref{tab:Correspondence}
and the observed average 
$\overline{n}$, with
an upper bound of 7 on $n$, since no more than
seven disks of radius 1.0 can touch a central disk of radius 1.5
(Section~\ref{sec:ModelII}).
The comparison of theory and simulations
(Fig.~\ref{fig:Estimated-and-measured}b) is less supporting than
that of the valences. 
Coordination numbers of 2, 5, 6 and 7 are much less
frequent than the predictions; 
coordination numbers of 3 and 4 are under-predicted
in their frequency.
\subsection{Discussion of Model~I results}\label{sec:discussion}
The theory captures features of the void valence probabilities,
but gives a rather poorer prediction of the distribution of coordination
numbers. 
Discrepancies result from a theory that addresses
topological disorder alone, but is uninformed by the geometric or
kinetic aspects of granular flow. 
That is, the theory is unburdened
by the reality that each vertex or face is a real particle or void,
having geometric character (size and shape) and an obligation to interact
with other vertices or faces while respecting kinematic (geometric),
equilibrium, and boundary constraints. 
These constraints are manifested in several ways.
Large coordination numbers $n$ are discouraged by
the geometric impossibility (or unlikelihood)
of fitting numerous particles around a central particle;
whereas, an $n=2$ is discouraged by a frictional limit
that can cause a central particle to ``squirt'' from
its two neighbors.
These two constraints are examined in the following section.
\par
In regard to the void valence distribution 
(Fig.~\ref{fig:Estimated-and-measured}a),
the occurrence of high-valence voids is 
discouraged by their inherent instability~\cite{Hunt:2010a}, since
large voids are short lived and readily collapse onto their inner
rattler particles;
whereas,
small voids ($l=3$) are discouraged by the geometric necessity that
sets of three particles be coordinated with
small interior angles (less than $\pi /2$, as evident in
Fig.~\ref{fig:Particle-graphs}b) while avoiding interior rattlers.
\par
One also notes the spatial patterning of void valences in
Fig.~\ref{fig:Particle-graphs}b:
voids of valence 3 and 4 are typically clustered to form ladder-like chains,
and voids of valence 6 and greater are also clustered near each other.
Examples of such chains and clusters are shaded orange and green
in Fig.~\ref{fig:Particle-graphs}b
(these patterns are more apparent when the periodic figure is multiply
tiled).
Stress transmission and deformation are also know to
be spatially organized into ``force chains'' and ``micro-bands''
\cite{Radjai:1996a,Kuhn:1999a,Azema:2007a}, and these patterns
affect the statistics of force and motion \cite{Radjai:1998a}
in a manner that is not yet fully understood \cite{Chakraborty:2010a}.
The spatial ordering of void valences
is certainly beyond 
Eqs.~(\ref{eq:Constraint2})--(\ref{eq:PastlDeltaM}).
\par
With its limitations, the model is not fully predictive, 
but even this primitive
view of granular flow as a solely topological process yields a reasonable
prediction 
of the distribution of void valences.
An entropy-based theory can, of course, 
be improved by supplying additional information
(pre-knowledge)
that further restricts the configuration space, and such
information, in the form of
geometric and equilibrium biases, 
is applied to the distribution of coordination numbers
in the next two sections.
\section{Geometric and kinetic constraints (Model~II)\label{sec:ModelII}}
\begin{figure}
  \centering
  \includegraphics{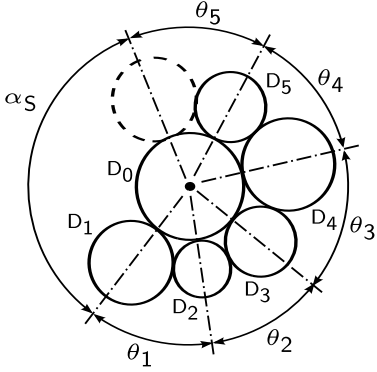}
  \caption{Available (excess) angular space $\alpha_{\tens{S}}$
           with five neighbors 
           $\tens{S}=\{ D_{1},D_{2},D_{3},D_{4},D_{5} \}$ 
           around particle $D_{0}$.
           The dashed disk is the image of $D_{1}$ that is used in
           defining $\theta_{5}$.}
  \label{fig:Alpha}
\end{figure}

Recognizing that geometric limitations can affect particle packings,
we now consider the ease (and, by extension, the likelihood) with which $n$
outer particles of sizes $\tens{S}=\{ D_{1}, D_{2}, \ldots, D_{n}\}$
can be packed around an inner particle of size $D_{0}$,
such that all particles of the ordered set $\tens{S}$ 
touch the inner particle (Fig.~\ref{fig:Alpha}).
Unlike the dense packing algorithm in\ \cite{Hihinashvili:2012a},
we permit loose packings in which gaps occur among the outer (shielding)
particles.
We evaluate the angle $\theta_{j}$ between the center of particle $j$ in
$\tens{S}$ and its neighbor $j+1$, 
when the two particles, $D_{j}$ and $D_{j+1}$, are temporarily
placed together and in contact with $D_{0}$
($\theta_{n}$ is the final angle between $D_{n}$
and $D_{1}$).
In some cases, it is not possible to pack the $n$ particles around $D_{0}$:
if the sum $\sum_{j=1}^{n}\theta_{j}$ exceeds $2\pi$, 
we assign a probability of zero for this combination of particles.
In other cases, the $n$ particles will fit with excess space.
We contend that the likelihood
of these arrangements is proportional to the excess angular space (gap)
$\alpha_{\tens{S}} \equiv 2\pi - \sum_{j=1}^{n}\theta_{j}$,
as in Fig.~\ref{fig:Alpha}.
The probability of encountering a particular arrangement
$\tens{S} = \{ D_{1}, D_{2}, \ldots, D_{n}\}$
around a central particle $D_{0}$ will depend on their
gap $\alpha_{\tens{S}}$ and on the distribution of particle sizes
in the entire assembly (i.e., the probabilities of the individual sizes
$D_{0},  D_{1}, D_{2}, \ldots$), as described below.
\par
\begin{figure}
  \centering
  \includegraphics{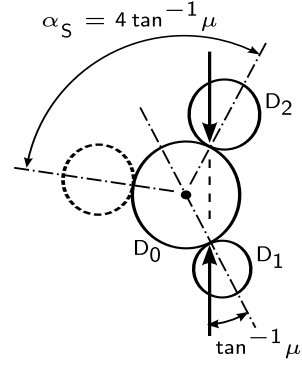}
  \caption{Available (excess) angular space 
           for two outer neighbors
           around inner particle $D_{0}$:
           $\alpha_{\tens{S}}=4\tan^{-1}\mu$.
           The dashed disk shows the range of stable
           locations of the disk $D_{2}$.}
  \label{fig:Squirt}
\end{figure}

Besides this geometric influence, a kinetic constraint applies 
in the case of only two outer particles, $n=2$,
due to the limiting friction coefficient $\mu$ (Fig.~\ref{fig:Squirt}).
Unless the exterior angle between the two outer particles is less
than $2\tan^{-1}\mu$, equilibrium can not be maintained, and the inner
particle will squirt from its two neighbors.
As a result, an angular range (gap) $\alpha_{\tens{S}}=4\tan^{-1}\mu$ 
is available for the arrangement of two particles around
an inner particle.
Within this available angular range, the mobilized friction can be
as small as zero (when the particles are diametrically opposed) and
as large as $\mu$ (when the particles are 
in the limiting arrangements shown in Fig.~\ref{fig:Squirt}),
an assumption that is consistent with experimental results\ \cite{Majmudar:2005a}.
\par
A poly-disperse assembly with 
$r$ different particle sizes admits $r^{n}$ possible
ordered sets $\tens{S}$ of cardinality $n$. 
We assume knowledge of the distribution of the $r$ sizes:
size $D_{k}$ comprises fraction $P_{D_{k}}$ of an entire assembly,
such that $\sum_{k=1}^{r}P_{D_{k}}=1$. 
We contend
the likelihood of encountering a particular ordered combination
$\tens{S}$ (of $n$ particles around an inner particle of size $D_{0}$) 
among all such combinations of cardinality $n$ is proportional
to three factors:
to the gap $\alpha_{\tens{S}}$ among the $n$ particles in $\tens{S}$,
to the probability $P_{D_{0}}$ of the inner size $D_{0}$, and to the
product of the $n$ probabilities of the outer sizes comprising $\tens{S}$:
\begin{equation}
  \alpha_{\tens{S}}\,
  P_{D_{0}}
  \prod_{k\in\tens{S}} P_{D_{k}}
  \label{eq:estimateIIpre0}
\end{equation}
If so,
the probability of encountering a particle with $n$ neighbors in
the entire assembly
is proportional to the sum $Q_{n}$ of all such $r^{n}$ combinations of 
$\tens{S}$ and all $r$ sizes $D_{0}$:
\begin{equation}
  Q_{n} = \sum_{\ell=1}^{r} 
             P_{D_{0,\ell}} 
          \sum_{j=1}^{r^{n}}
          \alpha_{\tens{S}_{j}}
          \prod_{k\in\tens{S}_{j}} P_{D_{k}}
  \label{eq:estimateIIpre}
\end{equation}
The sum of the probabilities for all coordination numbers, $n=2,3,\ldots$,
must equal one, so that the estimated probability of coordination
number $n$ is
\begin{equation}
  P_{n}^{\text{II}} = Q_{n} \left/ \sum_{n=2,3,\ldots} Q_{n}\right.
  \label{eq:estimateII}
\end{equation}
This equation gives
the probabilities of coordination numbers $n$ for the second ``II'' model.
Although no entropy principle is explicitly applied, we assume that
the probabilities in Eq.~(\ref{eq:estimateII}) are not
biased by factors other than geometric packing constraints and the kinetic,
frictional constraint.
Within these constraints, maximum entropy
is attained for each set of sizes $\tens{S}$ 
by a uniform
distribution of gaps $\alpha$ within the range $|\alpha | \leq \alpha_{\tens{S}}$.
\par
Anisotropy of particle arrangements is a dominant feature of granular
flow at the critical state\ \cite{Radjai:2012a,Kuhn:2010a}.
The model does not explicitly incorporate macro-scale geometric anisotropy,
but the model does permit anisotropy at the micro-scale:
for example, the particles $D_{1}$ and $D_{2}$ in
Fig.~\ref{fig:Squirt} are vertically aligned, 
and during vertical compression this arrangement
would be more likely than a horizontal alignment.
\subsection{Simulation results and Model~II}
\begin{figure}
  \centering
  \includegraphics{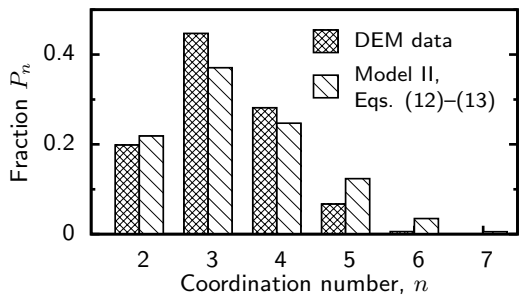}
  \caption{Estimated distribution of coordination numbers
           from model~II.}
  \label{fig:EstimatedII}
\end{figure}

With a bi-disperse assembly ($r=2$), the gaps and probabilities of all $r^{n}=2^{n}$
combinations of sizes can be computed in reasonable time.
The two particle sizes
sizes, $D_{a}=1$ and $D_{b}=1.5$, were in proportions
$P_{D_{a}}=2.25/3.25$ and $P_{D_{b}}=1/3.25$, such that both species
comprise the same total area.
Figure~\ref{fig:EstimatedII} 
shows the estimated probabilities $P_{n}^{\text{II}}$
for assemblies with $\mu=0.50$.
This figure compares the estimated distribution $P_{n}^{\text{II}}$
with the DEM simulations of 
dense granular flow at the critical state.
These DEM results are also shown in 
Fig.~\ref{fig:Estimated-and-measured}b.
The model~II yields a better representation of
the coordination number probabilities than model~I, 
capturing the general trend of the simulation results and
yielding a smaller fraction having $n=2$.
The model, however, over-estimates the average coordination number:
the average $\overline{n}$ of the simulations is $3.23$;
whereas, the estimated average is $3.40$.
\par
\begin{figure}
  \centering
  \includegraphics{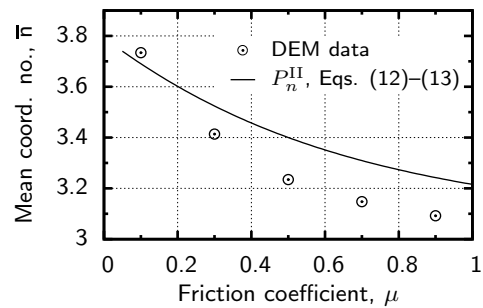}
  \caption{Friction coefficient $\mu$ between particles
           and the average coordination number:
           model~II estimates and simulation results.}
  \label{fig:EstimatedIImu}
\end{figure}

The arguments that were used in developing this method
would suggest that
the estimated probability of a 
particle having only two outer contacts is affected,
in part, by the friction coefficient $\mu$:
larger coefficients permit a greater range of orientations of two
outer particles (see Fig.~\ref{fig:Squirt}) 
and should increase the likelihood of
coordination number~2.
The average coordination number should, therefore, be smaller for
assemblies having a larger friction coefficient.
We conducted additional simulations with five different coefficients $\mu$.
These coefficients lead to different topological arrangements,
as expressed in different average coordination numbers $\overline{n}$
and in different distributions of $n$.
The average coordination numbers are shown in Fig.~\ref{fig:EstimatedIImu} and
are compared with the estimates of 
Eqs.~(\ref{eq:estimateIIpre})--(\ref{eq:estimateII}).
The decrease in $\overline{n}$ with increasing $\mu$ is consistent
with results in\ \cite{Shaebani:2012a}.
The estimates of the average coordination number $\overline{n}$
follow the trend of the simulation data, 
but $\overline{n}$ is over-predicted 
for all but the lowest friction coefficients, $\mu\approx 0.10$.
\par
Other simulations of moderately poly-disperse materials\ \cite{Shaebani:2012a}
have shown that the average deficit angle $\alpha_{\tens{S}}$ is about
the same for the smaller and larger particles in an assembly.
These results were reported for the initial, compacted state, 
prior to deviatoric loading.
At the critical state for the bi-disperse simulations, 
we measured average deficits 
of 162$^{\circ}$ and 172$^{\circ}$
for the small and large particles.
Eqs.~(\ref{eq:estimateIIpre0})--(\ref{eq:estimateII})
include all possible clustered arrangements $\tens{S}$,
with the probability of each arrangement being
naively weighted by its 
deficits $\alpha_{\tens{S}}$.
An estimate of the coordination number distribution
can be improved by including information (pre-knowledge)
of the average deficit for each central particle size.
In the next section,
we instead apply information of the average coordination number
$\overline{n}$ to improve the estimated distribution of $n$.
\section{Disorder with geometric and kinetic biases (Model~III)}\label{sec:ModelIII}
In this section, we develop an amalgam of the previous two approaches
by applying Kullback's minimum cross-entropy (MinXEnt) principle
\cite{Kullback:1951a,Kapur:1992a}.
To the author's knowledge, this principle has not yet been applied to granular
materials.
As with Jaynes' maximum entropy (MaxEnt, model~I) 
approach in Section~\ref{sec:ModelI},
the average $\overline{n}$ is applied as a rigid moment constraint
on the probability distribution $P_{n}$, 
but we also profess certain inclinations
of the probabilities in the form of ``\emph{a priori} estimates''
$q_{n}$.
The directed distance
$H^{\text{III}}_{n}(\vec{P}:\vec{q})$ between the two
distributions~--- from $\vec{q}$ to $\vec{P}$~--- is
the quasimetric
\begin{equation}
H^{\text{III}}_{n}(\vec{P}:\vec{q}) =
\sum_{n=2}^{\infty}
\sum_{\Delta M=1}^{n-1}
     P_{n,\Delta M}^{\text{III}}\ln\left(\frac{P_{n,\Delta M}^{\text{III}}}{q_{n}}\right)
\label{eq:HIII}
\end{equation}
In our case,
the probabilities $P_{n}^{\text{II}}$ derived in the previous
section from geometric and kinetic considerations are applied as the
\emph{a priori} estimates $q_{n}$ of coordination numbers $n$:
\begin{equation}
q_{n} = P_{n}^{\text{II}}
\label{eq:qn}
\end{equation}
\par
The MinXEnt principle calls for minimizing the distance $H^{\text{III}}_{n}$
subject to the applicable moment constraints, which will lead to a new
set of probabilities $P_{n,\Delta M}^{\text{III}}$.
As before,
the average coordination number $\overline{n}=\langle n \rangle$
and contact additions 
$\overline{\Delta M}=\langle \Delta M \rangle = \overline{n}/2$
are the moment constraints 
that apply to the problem of coordination numbers
(see Section~\ref{sec:ModelI}
and Table~\ref{tab:Correspondence}):
\begin{align}
    &\sum_{n=2}^{\infty}\sum_{\Delta M=1}^{n-1} n\, P_{n,\Delta M}^{\text{III}}
      = \overline{n} \label{eq:ConstraintIIIa} \\ 
    &\sum_{n=2}^{\infty}\sum_{\Delta M=1}^{n-1} 
      \Delta M\, P_{n,\Delta M}^{\text{III}}
      = \frac{\overline{n}}{2}
    \label{eq:ConstraintIIIb}
\end{align}
Minimizing (\ref{eq:HIII}) with respect to the 
$P_{n,\Delta M}^{\text{III}}$ subject to 
Eqs.~(\ref{eq:ConstraintIIIa}) and~(\ref{eq:ConstraintIIIb})
leads to the probabilities of the third ``III'' model:
\begin{gather}
  P_{n,\Delta M}^{\text{III}}=
  \frac{q_{n}}{Z_{n}^{\text{III}}\left(\lambda_{n,1},\lambda_{n,2}\right)}
  \exp\left(-\lambda_{n,1}n-\lambda_{n,2}\Delta M\right)
  \label{eq:PlDeltaMIII}\\
  Z_{n}^{\text{III}}\left(\lambda_{n,1},\lambda_{n,2}\right)=
  \sum_{n=2}^{\infty}\sum_{\Delta M=1}^{n-1}
  \! q_{n}
  \exp\left(-\lambda_{n,1}n-\lambda_{n,2}\Delta M\right)
  \label{eq:PartitionIII}
\end{gather}
with Lagrange multipliers, $\lambda_{n,1}$ and $\lambda_{n,2}$, that
satisfy Eqs.~(\ref{eq:ConstraintIIIa})
and~(\ref{eq:ConstraintIIIb}).
\par
The estimated probability $P_{n}^{\text{III}}$
of encountering coordination number $n$
is the marginal probability
\begin{equation}
P_{n}^{\text{III}} = \sum_{\Delta M=1}^{n-1}
P_{n,\Delta M}^{\text{III}}
\label{eq:PnIII}
\end{equation}
\par
This approach incorporates more information
than either of the previous methods:
it combines
the rigid moment constraints of the average coordination number $\overline{n}$
and the average added contacts $\overline{\Delta M}$, 
along with the geometric and kinetic inclinations $q_{n}$, 
which are intentional
biases that arise from the friction coefficient and distribution of particle
sizes.
If the inclinations $q_{n}$ were to yield, by themselves, the
same averages $\overline{n}$ and $\overline{\Delta M}$
(that is, if $\sum n\,q_{n}=\overline{n}$), 
then the moment constraints
of Eqs.~(\ref{eq:ConstraintIIIa}) and~(\ref{eq:ConstraintIIIb})
bring no further information beyond
that provided by the $q_{n}$.
In this case, the distribution $P_{n}^{\text{III}}$ will coincide with
the $P_{n}^{\text{II}}$ of Section~\ref{sec:ModelII}.
On the other hand, if the geometric and kinetic inclinations do not bias the
results (that is, if $q_2 = q_3 = \ldots$), then the inclinations
of Eq.~(\ref{eq:qn}) bring no further information beyond that
of Eqs.~(\ref{eq:ConstraintIIIa}) and~(\ref{eq:ConstraintIIIb}),
and the distribution $P_{n}^{\text{III}}$ will equal the $P_{n}^{\text{I}}$
of Section~\ref{sec:ModelI}.
(We note a subtle inclusion of such inclinations in that section,
when we disallowed coordination numbers greater than seven.  In a sense,
we had assumed $q_2 = q_3=\ldots =q_{7}=1/6$, $q_8=q_9=\ldots = 0$.)
\subsection{Simulation results and Model~III}
\begin{figure}
  \centering
  \includegraphics{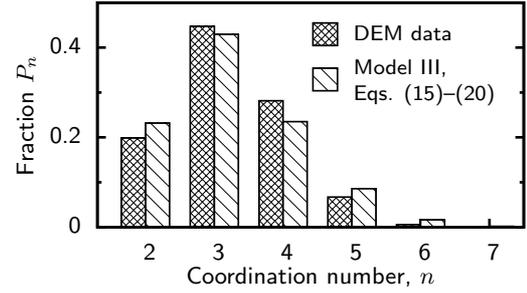}
  \caption{Estimated distribution of coordination numbers
           from model~III.}
  \label{fig:EstimatedIII}
\end{figure}

The distribution of coordination numbers predicted by model~III 
(Fig.~\ref{fig:EstimatedIII})
is fairly close to the DEM results and is better than those 
predicted with
models~I and~II (Figs.~\ref{fig:Estimated-and-measured}b 
and~\ref{fig:EstimatedII}).
As with model~I, which predicts a uniform distribution across
$\Delta M$ for each $l$ (Fig.~\ref{fig:Probabilities-DM}),
model~III predicts a uniform distribution
across $\Delta M$ for each $n$.
\par
All three estimates of coordination number distribution are summarized
in Table~\ref{tab:Compare},
which compares them with the DEM results.
In a recursion of the directed distance of Eq.~(\ref{eq:HIII}),
we use the following sum as the distance 
of an estimated distribution from the
measured DEM distribution $P_{n}^{\text{DEM}}$,
\begin{equation}
  \sum_{n=2,3,\ldots} P_{n}^{(\bullet )} 
  \ln \frac{P_{n}^{(\bullet )}}{P_{n}^{\text{DEM}}}
  \label{eq:distance2}
\end{equation}
where $P_{n}^{(\bullet )}$ is a particular estimate
($P_{n}^{\text{I}}$, $P_{n}^{\text{II}}$, or $P_{n}^{\text{III}}$).
\begin{table}
\caption{Distances between DEM results and estimates 
         of the coordination number distribution
         (Eq.~\ref{eq:distance2})}
\label{tab:Compare}
\centering
\begin{tabular}{lc}
\hline
Estimate & Distance \\
\hline
Model~I   & 0.268 \\
Model~II  & 0.077 \\
Model~III & 0.019 \\
\hline
\end{tabular}
\end{table}
In regard to the distribution of coordination numbers, the model~II,
which respects only geometric and kinetic information, yields a better
estimate than a model~I, 
which disregards such information and only considers topological
dispersion.
Model~III, which combines these factors, gives the best estimate.
\section{Conclusion}
When a granular material is slowly sheared from an initial rested condition,
the critical state is eventually attained at large shear strains,
and the bulk characteristics 
of this state --- density, fabric, and strength --- are insensitive to
the initial particle arrangement.
Because of this resilience, 
the critical state is used as a reference state
against which other conditions are compared 
(e.g., the jamming threshold in powder flows 
or with the so-called \emph{state parameter} 
in geomechanics \cite{Been:1991a}).
The critical state seems to be characterized
by maximum disorder, as simple disorder models are
shown to be
modestly successful in predicting the distributions of void valence
and coordination number.
Such disorder is moderated 
by a tendency toward spatial patterning
of the topological arrangement 
(Section~\ref{sec:discussion}) 
and by a bias toward certain
$(l_{i},\Delta M_{i})$ pairs in the particle graph 
(Fig.~\ref{fig:Probabilities-DM}).
We should also note that mono-disperse assemblies have a tendency
to crystallize into hexagonal arrangements of particles, although this
was not observed in the bi-disperse simulations of this study.
\par
Notwithstanding these tendencies for greater order,
the view of the critical state as a maximally disordered
condition could also be extended to other micro-scale characteristics,
such as contact forces, inter-particle motions, and contact orientation,
and these matters should be the focus of future study.  
The method should also be extended to three-dimensional assemblies,
although this poses new difficulties: in particular, a journaling
scheme for constructing a given granular topology by progressively
adding particles, contacts, and polygonal faces around
an existing arrangement to form new volume cells.
Finally, greater understanding should be sought for the
subtle tendency of greater order among the
occurrences of $\Delta M$
for each $n$ and $l$ (as in Fig.~\ref{fig:Probabilities-DM})
and for the effects of macro-scale anisotropy on the topological entropy.
These and other extensions of maximum disorder models will
lead to a richer understanding of critical state flow.
\begin{acknowledgements}
This work is dedicated to the memory of Prof. Colin B. Brown
(1929--2013), who made significant contributions 
to the understanding of granular entropy.
\end{acknowledgements}
%

\end{document}